\documentstyle[12pt]{article}%
\renewcommand{\title}[1]{\large\bf 
     #1\bigskip\medskip\\} 
\renewcommand{\author}[1]{\large #1\\ \smallskip}
\newcommand{\address}[1]{{\normalsize\it #1\\}\bigskip}

\newcommand{\be}{\begin{eqnarray}}
\newcommand{\ee}{\end{eqnarray}} 
\newcommand{\hs}[1]{\hspace*{#1cm}}
\newcommand{\vs}[1]{\vspace*{#1cm}}
\newcommand{\no}{\nonumber}
\newcommand{\wt}[6]{#1\mbox{\small
 $\left(\matrix{#5&#4\cr#2&#3\cr}\biggm|\mbox{$#6$}\right)$}}

\newcommand{\Km}[5]{{#1}\biggl(\!\matrix{&#3\vs{-0.3}\cr\!\!
  #2\hs{-0.3}\vs{-0.3}&\cr&#4}\!\!\biggm|\!\mbox{$#5$}\biggr)}
\newcommand{\Kp}[5]{{#1}\biggl(\matrix{#3\vs{-0.3}&\cr&
  \hs{-0.3}#2\vs{-0.3}\cr#4&}\!\!\biggm|\!\mbox{$#5$}\biggr)}
\newcommand{\half}{\mbox{$\textstyle {1 \over 2}$}}

\begin{document}
\begin{flushright} ANU Preprint MRR 073-95
\end{flushright}
\begin{center}
\title{Exact solution and surface critical behaviour\\ of open cyclic 
SOS lattice models} 
\author{Yu-Kui Zhou and Murray T. Batchelor}
\address{Department of Mathematics, School of Mathematical Sciences,\\
         Australian National University, Canberra ACT 0200, Australia}

\begin{abstract} 
We consider the $L$-state cyclic solid-on-solid lattice
models under a class of open boundary conditions. The integrable 
boundary face weights are obtained by solving the 
reflection equations. Functional relations for 
the fused transfer matrices are presented for 
both periodic and open boundary conditions. The 
eigen-spectra of the unfused transfer matrix is
obtained from the functional relations 
using the analytic Bethe ansatz.
For a special case of crossing parameter 
$\lambda=\pi/L$, the finite-size corrections to the 
eigen-spectra of the critical models are obtained, from which 
the corresponding conformal dimensions follow. 
The calculation of the surface free energy away from 
criticality yields two surface specific heat exponents,
$\alpha_s=2-L/2\ell$ and $\alpha_1=1-L/\ell$, where
$\ell=1,2,\cdots,L-1$ coprime to $L$. These results are in agreement with 
the scaling relations $\alpha_s=\alpha_b+\nu$ and 
$\alpha_1=\alpha_b-1$.
\end{abstract}
\end{center}
\bigskip
\hskip 1cm 27 Oct 95, to appear in J. Phys. A
\clearpage

\subsection{Introduction}\setcounter{equation}{0}
\setcounter{equation}{0}
Square lattice models in statistical mechanics with 
non-periodic boundary conditions have received intermittent
attention over the years (see, e.g. [1-6]). Up until quite 
recently the systematic study of the integrability of such 
non-periodic systems lagged well behind the study of the
corresponding periodic systems. It is well understood 
that models with periodic boundary conditions are 
integrable when their bulk/Boltzmann weights satisfy the 
Yang-Baxter equation \cite{Baxter}. Since Sklyanin's work 
\cite{Sklyanin}, we now understand that lattice 
models with open boundary conditions are integrable if in addition
the boundary weights satisfy the reflection equations 
\cite{Cherednik}. In particular, Sklyanin formulated the construction 
of commuting transfer matrices for the six-vertex model with 
open boundary conditions, from which the integrability is assured.
Recent lectures on subsequent developments can be found in 
\cite{MN,Kulish}.

Beyond their intrinsic mathematical interest, 
exactly solvable lattice models with open boundary conditions 
are attractive from the viewpoint of studying various  
surface critical phenomena [12-21]. 
Our motivation here is to study the surface critical behaviour 
of square lattice cyclic solid-on-solid (CSOS) 
models \cite{KuYa:88,PeSe:89}.
These models are face models in which the adjacency condition between
neighbouring heights is defined by the Dynkin 
diagram of the affine $A_{L-1}^{(1)}$ algebra.
The CSOS model has been well-studied for periodic boundary conditions.
The free energy, the local height probabilities and the
correlation length have all been evaluated, along with their
corresponding bulk critical exponents [22-25]. 
The complete operator content has been discussed in \cite{KiPe:89}
and the fusion procedure has been carried out in \cite{TDW:92,KNP:93}.
Some surface properties have been derived in \cite{SN}.  

The crossing or anisotropy parameter of the CSOS models is defined by
$\lambda = \ell \pi/L$, where $\ell=1,2,\cdots,L-1$ is coprime
to $L$ \cite{PeSe:89}. A special case of interest is $L=3$ and $\ell=2$ 
which is related to Baxter's three colourings of the square 
lattice \cite{Baxter:70,Baxter,TrSc:86,PeSe:89}.

The layout of this paper is the following. In section 2 the CSOS
models with both periodic and open boundary conditions are 
described. We solve the reflection equations for the  
boundary face weights. The functional relations of the fused 
transfer matrices are also presented. In section~\ref{solutions}
the eigen-spectra of the unfused transfer matrix is extracted 
from the functional relations following the analytic 
Bethe ansatz method. The finite-size corrections to the transfer
matrix eigen-spectra at criticality are obtained for a special
value of the crossing parameter. In section~\ref{surface} the 
free energy of the open boundary models is shown to 
satisfy a unitarity relation. We solve the unitarity relation 
following the inversion relation method \cite{Baxter,Baxter:82}.
From the singular part of the free energy we obtain two 
surface specific heat exponents in agreement with scaling 
predictions. We conclude with a brief discussion.

\subsection{CSOS models}\setcounter{equation}{0}
\setcounter{equation}{0}\label{models}

The CSOS lattice models \cite{KuYa:88,PeSe:89} are
a family of $L$-state face models \cite{Baxter} 
built on the affine $A_{L-1}^{(1)}$ Dynkin diagram. States at 
adjacent sites of the
square lattice must be adjacent on the Dynkin diagram.  
The cyclic nature of the heights distinguishes the CSOS model
from the corresponding RSOS model \cite{ABF} built on the $A_L$ Dynkin 
diagram.   

\subsubsection{Bulk face weights}

The allowed, or non-zero, face weights of the CSOS models are 
given by \cite{KuYa:88,PeSe:89}
\addtolength{\jot}{2mm}
\be
\wt Wa{a\pm 1}a{a\mp 1}u & = & {\vartheta_1(\lambda-u)
       \over \vartheta_1(\lambda)} \no \\
\wt W{a\pm 1}a{a\mp 1}au & = &
\left[\frac{\vartheta_4(w_{a-1})\vartheta_4(w_{a+1})}{
  \vartheta_4^2(w_a)}\right]^{1/2}
 \frac{\vartheta_1({u})}{\vartheta_1({\lambda})}  \\
\wt W{a\pm 1}a{a\pm 1}au & = &
           \frac{\vartheta_4({w_a\pm u})}{\vartheta_4(w_a)} \no
\label{bface}\ee
where  $w_a=a\lambda+w_0$. The height $a=1,2,\cdots,L$ and 
$0<w_0<\pi$ is a free parameter. The crossing parameter 
$\lambda$ is 
given by $\lambda=\ell\pi/L$, where $\ell=1,2,\cdots,L-1$ is coprime 
to $L$ and $L>2$. The elliptic 
functions $\vartheta_1({u})$, $\vartheta_4({u})$ are standard 
theta functions of nome $p$ 
\be
\vartheta_1(u)&=&\vartheta_1(u,p)=2p^{1/4}\sin u\:
  \prod_{n=1}^{\infty} \left(1-2p^{2n}\cos
   2u+p^{4n}\right)\left(1-p^{2n}\right)\label{theta1}  \\
\vartheta_4(u)&=&\vartheta_4(u,p)=\prod_{n=1}^{\infty}\left(
 1-2p^{2n-1}\cos2u+p^{4n-2}\right)\left(1-p^{2n}\right)
  \label{theta4}
\ee
where $0<p<1$ with $p=0$ at criticality.

These face weights satisfy the star-triangle equation
\be
\sum_g\wt Wabgfu\wt {W}fgdev\wt {W}gbcd{v\!-\!u} \no \\
=\sum_g\wt {W}fage{v\!-\!u}\wt {W}abcgv\wt Wgcdeu 
\label{YBE} 
\ee
inversion/unitarity relations
\be
\sum_{g} \wt Wabgdu\wt Wgbcd{-u} =\rho(u)\delta_{a,c}
\ee
and the crossing unitarity relations
\be
\sum_{g} \wt Wdabg{\lambda-u} \wt
Wdgbc{\lambda+u} {\vartheta_4(w_{a})\vartheta_4(w_{g})
\over \vartheta_4(w_{d})\vartheta_4(w_{b})}=\rho(u)\delta_{a,c}
\ee
where $\rho(u)=\vartheta_1({\lambda-u})\vartheta_1({\lambda+u})/
\vartheta_1^2({\lambda})$.

\subsubsection{Periodic boundaries}

There is a hierarchy of commuting families of transfer 
matrices constructed by the fusion procedure on the CSOS 
models under periodic boundary conditions. Let  
$\wt {W_{m\times n} }{a}{b}{c\vspace*{-0.1cm}}du$ be the fused face weights
with fusion level $m$ and $n$ in the vertical and
horizontal directions, respectively \cite{TDW:92,KNP:93}. 
Then the fused transfer 
matrices $\mbox{\boldmath $T$}^{(m,n)}(u)$ are defined with elements
\begin{eqnarray}
\langle\mbox{\boldmath $a$}|\mbox{\boldmath $T$}^{(m,n)}(u)|
       \mbox{\boldmath $b$}\rangle =
\prod_{j=1}^N \wt {W_{m\times n}}{b_j}{b_{j+1}}{a_{j+1}}{a_j}u
\end{eqnarray}
with $a_{N+1}=a_1$ and $b_{N+1}=b_1$ where $N$ is the number
of faces in a row of the lattice. By construction the fused face weights 
satisfy the star-triangle equations, resulting in  
the commutation relations
\begin{equation}
[\mbox{\boldmath $T$}^{(m,n)}(u),
   \mbox{\boldmath $T$}^{(m,\overline{n})}(v)] = 0 .
\label{eq:rowcommute1}
\end{equation}

These fused transfer matrices satisfy groups of functional
relations, which can be easily proved by fusion. Let us define 
\be
\mbox{\boldmath $T$}^{(n)}_k&=&
   {\mbox{\boldmath $T$}}_{(m,n)}(u+k\lambda)           \no \\ 
\mbox{\boldmath $T$}^{(n)}&=&0   
   \hspace{0.5cm} \mbox{if $n<0$ or $m<0$}  \\
\mbox{\boldmath $T$}^{(0)}&=&{\bf I}      \no 
\label{s} 
\ee
along with the function
\be
 f^m_n&=&\prod_{j=0}^{m-1}\rho^N(u-j\lambda+n\lambda).
\ee
Then for each $m=1,2,\cdots$ the functional relations are
\be
&&\hspace{0.5cm}\mbox{\boldmath $T$}^{(n)}_0
  \mbox{\boldmath $T$}^{(1)}_n= 
   \mbox{\boldmath $T$}^{(n+1)}_0  + 
  f^m_{n-1}\mbox{\boldmath $T$}^{(n-1)}_0  
\hs{0.5}\mbox{$n=1,2,\cdots$}\label{fr1} 
\ee

The unfused models of interest here are recovered by setting
the fusion level to $n=m=1$.  Disregarding finite-size corrections, 
the bulk free energy in this case satisfies 
\be
T(u) T(u+\lambda)=f^1_{0}. 
\ee
This is the unitarity relation for periodic boundary conditions.

\subsubsection{Boundary face weights}

Integrable models with open boundary conditions are defined by
both the bulk and the boundary face weights.
The latter are represented by three heights interacting round
a triangular face \cite{Kulish,FHS:95,BOP:95,AK:95,Zhou:95b}.
For the CSOS models,
\be
\Km {K}acbu=0 \hspace*{0.5cm}
  \mbox{unless $|a-b|=1,L-1$ and $|a-c|=1,L-1$}
\ee
which satisfy the boundary version of the star-triangle equation 
(reflection equations)
\be
&&\sum_{f,g}{\wt Wgcba{u-v}}{\Km {K}gcf{u;\xi}}{\wt
  Wdfga{u+v}}{\Km {K}dfe{v;\xi}} 
  \no\\
&&\hspace{0.5cm} =\sum_{f,g} {\Km {K}bcf{v;\xi}}{
   \wt Wgfba{u+v}}{
    \Km {K}gfe{u;\xi}
     }{\wt Wdega{u-v}} \label{BYBE}
\ee
In general there may be more arbitary parameters 
than $\xi$. Inserting the CSOS bulk face weights 
(\ref{bface}) into the reflection equations and making use
of the elliptic function identity
\be
\vartheta_1(x+y) \vartheta_1(x-y) \vartheta_4(w+v) \vartheta_4(w-v) -
\vartheta_1(v+y) \vartheta_1(v-y) \vartheta_4(w+x) \vartheta_4(w-x) \no\\
=
\vartheta_1(x+v) \vartheta_1(x-v) \vartheta_4(w+y) \vartheta_4(w-y)
\hs{4}
\label{id}
\ee
we find the following CSOS boundary face weights
\be
\Km {K}{1}{L}{L}{u;\xi}\;=\;{\vartheta_1[\xi+u]
  \vartheta_4[u-(w_L+\xi)]\over 
  \vartheta_1^2(\lambda)}   \label{K1}\\
\Km {K}{L}{1}{1}{u;\xi}\;=\;{\vartheta_1[\xi-u]
  \vartheta_4[u+(w_1+\xi)]\over 
  \vartheta_1^2(\lambda)}   \\
\Km {K}{a}{t}{b}{u;\xi}\;=\;{\vartheta_1[\xi+(a-t)u]
  \vartheta_4[u-(a-t)(w_b+\xi)]\over 
  \vartheta_1^2(\lambda)}\delta_{b,t}.\label{K}
\ee
The identity (\ref{id}) also plays a role in establishing 
the integrability of the bulk weights \cite{PeSe:89}.

It is obvious that the boundary face weights satisfy the
crossing symmetry
\be
\sum_{c}\sqrt{\vartheta_4(w_c)\over\vartheta_4(w_a)}
\wt Wabcd{2u+\lambda} \Km Kceb{u+\lambda} = {\vartheta_1(
 2u+2\lambda)\over \vartheta_1(\lambda)}    \Km Kaeb{-u}\;.
\label{Kcrossing}\ee

\subsubsection{Fusion results}

The fused transfer matrices $\mbox{\boldmath $T$}^{(m,n)}(u)$
of the open boundary CSOS models are defined by
 the following elements
\be
\langle\mbox{\boldmath $a$}|\mbox{\boldmath $T$}^{(m,n)}(u)| 
  \mbox{\boldmath $b$}\rangle =\sum_{\{c_0,\cdots,c_N\}}\Kp 
 {K_+^{(n)}}{c_0}{a_0}{b_0}u\Km {K_-^{(n)}}{c_N}{
   a_N}{b_N}u\times \hspace{1cm} \no \\  
  \prod_{k=0}^{N-1}\biggl[\wt {W_{(n\times m)}}{b_k}{b_{k+1}
   }{c_{k+1}}{c_k}{u}\wt 
 {W_{(m\times n)}}{c_{k+1}}{a_{k+1}}{a_k}{c_k}{
   u+n\lambda-\lambda} \biggl]  \label{fopenT}
\ee
where the right boundary fused face weights $K_-^{(n)}$ are 
given by fusing 
\be
\Km {K_-}{a}{t}{b}u =\Km {K}{a}{t}{b}{u;\xi_-}
\ee
while the left boundary fused face weights $K_+^{(n)}$ are 
given by fusing 
\be
\Kp {K_+}{a}{t}{b}u =\Km
  {K}{a}{t}{b}{-u+\lambda;\xi_+}\sqrt{\vartheta^2_4(w_{a})
     \over  \vartheta_4(w_{t})\vartheta_4(w_{b})}.
\ee
These fused transfer matrices form commuting families, 
\be
\left[\mbox{\boldmath $T$}^{(m,n)}(u)\; ,
   \; \mbox{\boldmath $T$}^{(m,n^\prime)}(v)\;\right]=0\;.
 \label{fTT}
\ee
Like its periodic counterpart, the unfused transfer matrix 
is recovered by setting the fusion levels to $m=n=1$, i.e., 
$\mbox{\boldmath $T$}(u)=\mbox{\boldmath $T$}^{(1,1)}(u)$.

The fusion procedure for face models with open boundary
conditions has been demonstrated elsewhere \cite{BOP:95,Zhou:95b}
and thus we do not repeat the details here. However, it is  
worthwhile to write down the functional relations satisfied by the fused
CSOS transfer matrices. We find 
\be
\mbox{\boldmath $T$}^{(m,n)}_k&=&
   {\mbox{\boldmath $T$}}_{(m,n)}(u+k\lambda)           \no \\ 
\mbox{\boldmath $T$}^{(m,n)}&=&0   
       \hspace{0.5cm} \mbox{if $n<0$ or $m<0$}  \\
\mbox{\boldmath $T$}^{(m,0)}&=&{\bf I} 
\ee
where now the auxiliary function $f^m_n$ is determined by
\be         
 f^m_n&=&{\omega^-(u+n\lambda)\omega^+(u+n\lambda)\over
   \rho(2u+2n\lambda)}
  \prod_{j=0}^{m-1}\rho^N(u-j\lambda+n\lambda)
              \rho^N(u+j\lambda+n\lambda)\;. \label{fmn} 
\ee
The boundaries contribute the diagonal matrix factors 
$\omega^-(u)$ and $\omega^+(u)$, with
\be
\omega^-_{c_r,c_r}(u)=\sum_{a,b}\sqrt{\vartheta_4(w_b)\over 
    \vartheta_4(w_{c_r-1})}\wt 
  W{c_r\!-\!1\!\!}ab{c_r\!\!}{2u\!+\!\lambda}\!
 \Km {K_-}b{c_r}a{u\!+\!\lambda}\!\Km {K_-}{c_r\!-\!1}a{c_r}u \\
\!\omega^+_{c_l,c_l}(u)= \!\sum_{a,d}\!\sqrt{\!
    \vartheta_4^2(w_a)
    \vartheta_4^2(w_{c_l})\over \vartheta_4^3(w_{{c_l}-1})
   \vartheta_4(w_d)}\!\wt {W\!\!}{\!\!d\hs{-0.2}}{
  a\!\!}{{c_l}\!-\!1\!\!}{\!\!{c_l}\hs{-0.2}}{
  \!\!\lambda\!\!-\!\!2u}  \Kp {K_+}{{c_l}\!\!-\!\!1}{
   c_l}a{u\!\!+\!\!\lambda}\Kp {K_+}da{c_l}u. 
\ee
Here height $c_r$ ($c_l$) is located on the right (left) boundary.
The matrix functions $\omega^\pm(u)$ are simplified under the crossing 
symmetry (\ref{Kcrossing}), with
\be
\omega^\mp_{c,c}(u)={\vartheta_1({2\lambda\pm 2u})\over
  \vartheta_1({\lambda})}\sum_{a} \Km
  {K}{c\!-\!1}ca{\mp u;\xi_\mp}\Km {K}{c\!-\!1}ac{\pm u;\xi_\mp}.
\ee
Then the functional relations have similar forms to the
periodic boundary case, namely 
\be
&&\hspace{0.5cm}\mbox{\boldmath $T$}^{(m,n)}_0
   \mbox{\boldmath $T$}^{(m,1)}_n= 
   \mbox{\boldmath $T$}^{(m,n+1)}_0  + 
    f^m_{n-1}\mbox{\boldmath $T$}^{(m,n-1)}_0
     \hs{0.5}m,n\ge 0\;.\label{fr2} 
\ee

In this way, again after dropping the finite-size corrections,
we arrive at the unitarity relation
\be
{T}(u){T}(u+\lambda)=f(u)\;, \label{unit}
\ee
where the function $f$ is given by $f(u)=f^1_0(u)\rho(2u)
\vartheta_1^2({\lambda})/\vartheta_1^2({2\lambda})$ after
appropriate renormalization of the free energies.
This relation is sufficient to determine both the bulk and
surface free energies \cite{Zhou:95b}.

\subsection{Exact Bethe ansatz solution}\setcounter{equation}{0}
\setcounter{equation}{0}\label{solutions}

Sklyanin presented the algebraic Bethe ansatz solution of the six-vertex
model \cite{Sklyanin} or spin-\half \ XXZ chain \cite{ABBBQ:87} 
with open boundary conditions. Unfortunately, the
generalization of the algebraic Bethe ansatz to treat other 
integrable open boundary models has not made much progress,
in particular, for models in which the arrow or spin reversal symmetry
is broken. However, when such symmetry holds,  
the Bethe ansatz solutions of many integrable open boundary models 
have been obtained (see, e.g. \cite{MN,YuBa:95a,AMN:95}). Here we  
show that the analytic ansatz method \cite{Reshetikhin:83} can be applied 
to find the transfer matrix eigen-spectra of the CSOS models.
This method has also been applied with success to the CSOS models with
periodic boundary conditions \cite{PeBa:90}.

\subsubsection{Bethe ansatz solution}
Consider the transfer matrix 
$\wt {\mbox{\boldmath $T$}}{b_0}{b_N}{a_N}{a_0}u$, which  is the
transfer matrix $\mbox{\boldmath $T$}^{(1,1)}_0$ with fixed heights 
${b_N},{a_N}$ along the right boundary and $b_0,a_0$ along the
left boundary. For the solutions (\ref{K1})-(\ref{K}) the 
transfer matrix is nonzero only for $b_0=a_0$ and ${b_N=a_N}$.
Suppose $\mbox{\boldmath $T$}_{b,d}(u)
=\wt {\mbox{\boldmath $T$}}{b}{d}{d}{b}{u-\lambda}$.
Let us consider the following ansatz
\be
T_{b,d}(u)&=&\Km K{d-1}dd{u;\xi_-}\Km K{b+1}bb{u;\xi_+}
 {\vartheta_1({2u-2\lambda})\over\vartheta_1({2u-\lambda})} \no \\
 &&\hspace{2cm}\times\vartheta_1^{2N}(\lambda-u)
  \prod_{j=1}^M{\vartheta_1(u+\lambda+u_j)\vartheta_1(u-u_j)\over
               \vartheta_1(u-\lambda-u_j)\vartheta_1(u+u_j)}\no\\
&& +\Km K{d+1}dd{u-\lambda;\xi_-}\Km K{b-1}bb{u-\lambda;\xi_+}
 {\vartheta_1({2u}) \over\vartheta_1(2u-\lambda)}\no \\
&&\hspace{2cm}\times \vartheta_1^{2N}(u)
\prod_{j=1}^M{\vartheta_1(u-\lambda+u_j)\vartheta_1(u-2\lambda-u_j)
           \over\vartheta_1(u-\lambda-u_j)\vartheta_1(u+u_j)}
\label{eig}\ee
for the eigen spectra of the transfer matrix 
$\mbox{\boldmath $T$}_{b,d}(u)$. 
We can check that the above ansatz satisfies the functional relation 
(\ref{fr2}) if the parameters $u_j$ satisfy
\be
T_{b,d}(u_k)=0\hspace{0.6cm}k=1,2,\cdots,M. \label{BAE}
\ee
As a result, the ansatz (\ref{eig}) should give the eigen-spectra
of the CSOS transfer matrix, with (\ref{BAE}) as the related Bethe 
ansatz equations.  The eigen-spectra of the fused transfer matrices can 
been written down according to the fusion results 
from (\ref{eig}) and (\ref{BAE}).
 
\subsubsection{Finite-size corrections}

At criticality the CSOS models are equivalent to the six-vertex
model. The transfer matrix $\mbox{\boldmath $T$}_{b,d}(u)$ is then
independent of the boundary heights (as are the fused transfer 
matrices) and the Bethe ansatz solution (\ref{eig})-(\ref{BAE}) 
is of the same form as the solution of the open boundary 
six-vertex model \cite{Sklyanin,YuBa:95a,BOSY:95}. Now the 
finite-size corrections to the fused transfer matrices of the
$U_q(sl_2)$-invariant six-vertex model have been derived in \cite{Zhou:95a}.
Thus by a similar calculation we obtain
the finite-size corrections to the fused  transfer
matrices of the critical CSOS models. 

For simplicity we consider the limit 
$\xi_\pm\to i\infty$ with $\lambda=\pi/L$ and $N$ even. The fused CSOS 
transfer matrix eigenvalues then behave like
\be
\log T^{(p,p)}(u) 
&=&-2N f_b(u)-f_s(u)+{\pi\over 12N}\left(c-24\Delta_{1,\nu,r}\right)\sin(Lu)
 +{ o\!}\left({1\over N}\right).
\label{finite}
\ee 
The central charge is given by 
\be
c&=&{3p\over p+2}-{6p\over L(L-p)}\;  \label{c} 
\ee
where $p=1,2,\cdots,L-2$ labels the fusion level.
The conformal weights are given by
\be
\Delta_{1,\nu,r}\;=\;{\left( L-(L-p)r\right)^2-p^2\over 4Lp(L-p)} +
    {\nu(p-\nu)\over 2p(p+2)} \label{Kac}
\ee
with $\nu$ a unique integer determined by
\be 
\nu=r-1-p\lfloor{r-1\over p}\rfloor  \;. 
\ee
and $r=1,3,\cdots\le L-2$. Here $\lfloor{x}\rfloor$ is the largest
integer part less than or equal to $x$. The functions $f_b(u)$ and
$f_s(u)$ are, respectively, the bulk free energy and surface free 
energy of the critical models, which are also calculated for the 
off-critical models in the next section. They are not given 
explicitly here.

\subsection{Surface free energy and critical exponents} 
\setcounter{equation}{0}\label{surface}

The bulk and surface free energies of the CSOS models
can be found from the unitarity relation (\ref{unit}) with certain 
analyticity assumptions, as has been shown in the study of the
eight-vertex model \cite{Baxter,Baxter:82,BaZh:95}. 

The unitarity relation (\ref{unit})  
combines the inversion relation and crossing symmetries of 
the local bulk and boundary face weights. We can separate
the contributions from the bulk and surface free energies 
in this relation \cite{Zhou:95b}. Let $T(u)=T_b(u) T_s(u)$ be the 
eigenvalues of the transfer matrix $\mbox{\boldmath $T$}(u)$. Define 
$T_b = \kappa_b^{2 N}$ 
and $T_s = \kappa_s$, then the free energies are defined by 
$f_b(u)=-\log \kappa_b(u)$ and $f_s(u) = -\log \kappa_s(u)$.
We have
\be
\kappa_b(u)\kappa_b(u+\lambda)&=&
  {\vartheta_1(\lambda-u)\vartheta_1(\lambda+u)
    \over \vartheta_1(\lambda)\vartheta_1(\lambda)} \label{inv-b34}
\ee
for the bulk and 
\be
&& \hspace{-1.2cm}
\kappa_s(u)\kappa_s(u+\lambda)=
  {\vartheta_1({2\lambda+2u})\vartheta_1({2\lambda-2u})
  \over\vartheta_1^2({2\lambda})} \label{uni}\\  
&&\hspace{-0.5cm}\times  \Km {K}{d\!-\!1}dd{-u;\xi_-}
   \Km {K}{d\!-\!1}dd{u;\xi_-} 
  \Km {K}{b\!-\!1}bb{-u;\xi_+}
   \Km {K}{b\!-\!1}bb{u;\xi_+}  \no
\ee
for the surface. Here height $d$ ($b$) is located on the right 
(left) boundary. 

To solve the unitarity relations it is convenient to introduce
the new variables
\be
x=e^{-\pi\lambda/\epsilon}, &
w=e^{-2\pi u/\epsilon},&
q = e^{-\pi^2/\epsilon} \no \\
 v_a= e^{-\pi w_a/\epsilon},& 
 v_\pm=e^{-\pi\xi_\pm/\epsilon}, &p=e^{-\epsilon} 
\ee
along with the conjugate modulus transformation of
the theta functions,
\be
\vartheta_1(u,e^{-\epsilon})&=&\rho(u,\epsilon) 
     E\left(e^{-2\pi u/\epsilon},
        e^{-2\pi^2/\epsilon}\right) \\
\vartheta_4(u,e^{-\epsilon})&=&\rho(u,\epsilon) 
     E\left(-e^{-2\pi u/\epsilon},
        e^{-2\pi^2/\epsilon}\right). 
\ee
The factor $\rho(u,\epsilon)$ is harmless and will be 
disregarded, while
\be
E(z,x)=\prod_{n=1}^\infty(1-x^{n-1}z)(1-x^{n} z^{-1})(1-x^n).
\ee

Suppose that $\kappa_b(w)$ is analytic and nonzero in the annulus 
$x^2\le w\le 1$, we can Laurent expand $f_b(w)$ as 
$\log\kappa_b(w)=\sum_{n=-\infty}^{\infty} c_n w^n$.
Then inserting the series expansion 
into the logarithm of both sides of (\ref{inv-b34})
and equating coefficients in powers of $w$ gives \cite{PeSe:89}
\be
f_b(w,p)=-\sum_{n=1}^{\infty}
{(x^{2n}+q^{2n}x^{-2n})(1-w^n)(1-x^{2n}w^{-n})\over
          n(1+x^{2n})(1-q^{2n})}. \label{bfree}
\ee
Similarly, taking the Laurent expansion $\log\kappa_s(w)=
 \sum_{n=-\infty}^{\infty} c_n w^n$ and solving the
unitarity relation (\ref{uni}) yields
\be
f_s(w,\xi_\pm,p) 
&=&\sum_{n=1}^{\infty}
{(-1)^n(v_+^{2n}v_b^{2n}+q^{2n}v_+^{-2n}v_b^{-2n}+
   v_-^{2n}v_d^{2n}+q^{2n}v_-^{-2n}v_d^{-2n})
   (w^n+x^{2n}w^{-n})\over    n(1+x^{2n})(1-q^{2n})} \no \\
&&\hs{0.4}+\;\sum_{n=1}^{\infty}
{(v_+^{2n}+q^{2n}v_+^{-2n}+v_-^{2n}+q^{2n}v_-^{-2n})
   (w^n+x^{2n}w^{-n})\over    n(1+x^{2n})(1-q^{2n})} \no \\
&&\hs{0.4}-\;\sum_{n=}^{\infty}{(x^{4n}+q^{2n}x^{-4n})
   (1-w^{2n})(1-x^{4n}w^{-2n})\over   n(1+x^{4n})(1-q^{2n})}\no \\
&&\hs{0.4}-\;\sum_{n=1}^{\infty}{4(x^{2n}+q^{2n}x^{-2n})
   \over  n(1-q^{2n})} .
\label{sfree}
\ee
The surface free energy is explicitly dependent on the boundary 
heights $b,d$ and $\xi_\pm$.

The specific heat critical exponents may be obtained
from the leading order singularity of the free energies.
In practice, the singular behaviour is extracted by means of
the Poisson summation formula \cite{Baxter}. For the bulk free 
energy it follows that \cite{KuYa:88,PeSe:89} 
\be
f_b(w,p)\sim p^{\pi/\lambda}\quad\mbox{as}\quad p\to 0. 
\ee
When $\ell=1$ and $L$ is even there is a multiplicative $\log p$ factor. 
It follows that the bulk specific heat exponent of the CSOS models is  
$\alpha_b=2-{\pi\over\lambda}$ where we recall that $\lambda=\ell\pi/L$.
The same idea can be applied to find the surface specific heat 
exponents 
from the surface energy $f_s(w,\xi_\pm,p)$.
Following \cite{bin,ws} and the treatment of the eight-vertex model 
\cite{BaZh:95} we define the excess internal energy $e_s$, 
\be
e_s(p)\sim {\partial f_s(w,\xi_\pm,p)\over\partial p}+e_1(p),
\ee
where $e_1(p)$ is called as the local internal energy
in the surface layer, which is the correction energy  to the 
surface internal energy $e_s(p)$, given by
\be
e_1(p)\sim {\partial f_s(w,\xi_\pm,p)\over\partial \xi_\pm}\;.
\ee
The free parameter $\xi_\pm$ appearing in the boundary face weights
can be interpretted as a surface coupling. This has been explicitly 
shown in the study of the eight-vertex model \cite{BaZh:95}.
The corresponding specific heats are defined by
\be
C_s \sim {\partial e_s\over\partial p}\;,\hs{0.4}
C_1 \sim {\partial e_1\over\partial p}\;.
\ee

Application of the Poisson summation formula to  
$f_s(w,\xi_\pm,p)$ yields 
\be
e_s(p)&\sim& p^{{\pi\over 2\lambda}-1}\\
e_1(p)&\sim& p^{\pi\over \lambda}
\ee
as $p\to 0$, with a similar $\log p$ correction factor as for 
the bulk case. The surface specific heat exponents of the
CSOS models follow as 
\be
\alpha_s=2-{\pi\over 2\lambda}\hs{0.4}{\rm and}\hs{0.4} 
\alpha_1=1-{\pi\over \lambda}.
\ee
For the three colouring problem ($L=3$ and $\ell=2$) we thus
obtain the values $\alpha_s={5\over 4}$ and
$\alpha_1=-{1\over 2}$.

Recalling the bulk exponent $\alpha_b=2-{\pi\over\lambda}$ 
\cite{KuYa:88,PeSe:89} 
and the correlation length exponent $\nu={\pi\over 2\lambda}$ 
\cite{PeBa:90} we are thus able to provide a further  
significant test of the scaling relations
\cite{bin,ws}
\be
\alpha_s=\alpha_b+\nu\hs{0.4}{\rm and}\hs{0.4}
 \alpha_1=\alpha_b-1\;.
\ee

\subsection{Discussion}

In this paper we have derived exact results for the critical surface
properties of the CSOS lattice models. They 
can be generalized to the fused CSOS models.
In this and related work \cite{Zhou:95b,BaZh:95,ZhBa:95} the crossing 
unitarity relation plays a key role
in deriving the surface free energy away from criticality. 
The CSOS lattice models with fixed boundary conditions on the
square lattice with diagonal orientation can be treated in a
similar manner by incorporating inhomogeneities into the bulk
face weights and taking appropriate values of the inhomogeneities
and the boundary couplings $\xi_\pm$, as has been
explicitly demonstrated for the ABF RSOS models \cite{ZhBa:95}.
However, we do not pursue this direction here as 
the change in lattice orientation does not effect the
critical exponents.

The excess surface critical exponent $\alpha_s$  
has been obtained from the singular
leading term of the excess internal energy $e_s$. It turns out that
the singular leading term does not depend on the boundary face 
weights and thus $\alpha_s$ is
independent of the details of the boundary weights. This behaviour
has been already seen in the study of ABF model \cite{ZhBa:95}.

Other models of immediate interest are the 
dilute $A_L$ models \cite{WNS,Roche} which can be obtained from 
Kuniba's $A_2^{(2)}$ face model \cite{Kuniba} under restriction.
The boundary face weights have been found for these models and the surface
critical properties can thus be studied in a similar way
\cite{BFZ:95}.

\subsection*{Acknowledgements} This research has been supported 
by the Australian Research Council. 


\end{document}